\documentclass{acm_proc_article-sp}

\usepackage{amsmath,amssymb,color,bbm,ifthen,graphicx}
\usepackage[latin1]{inputenc}
\usepackage[mathcal]{euscript}
\usepackage[english]{babel}

\newtheorem{theo}{Theorem}
\newtheorem{lemma}{Lemma}

\newtheorem{coro}{Corollary}
\newtheorem{assum}{Assumption}

\newcounter{hypA}

\newcounter{hyp}

\newcommand{\un}{\ensuremath{\mathbbm{1}}}

\newcommand{\bs}{\boldsymbol}

\newcommand{\sW}{{\mathsf W}}
\newcommand{\sd}{{\mathsf d}}

\newcommand{\cP}{{\mathcal P}}

\newcommand{\cL}{{\mathcal L}}

\newcommand{\cF}{{\mathcal F}}
\newcommand{\cN}{{\mathcal N}}

\newcommand{\cD}{{\mathcal D}}
\newcommand{\cV}{{\cal V}}
\newcommand{\cG}{{\cal G}}
\newcommand{\cW}{{\cal W}}
\newcommand{\cE}{{\cal E}}

\newcommand{\bP}{{\mathbb P}}
\newcommand{\bR}{{\mathbb R}}

\newcommand{\bE}{{\mathbb E}}

\newcommand{\bN}{{\mathbb N}}

\newcommand{\dG}{\partial G}

\newcommand{\la}{\langle}
\newcommand{\ra}{\rangle}

\newcommand{\bth}{{\bs \theta}}
\newcommand{\ath}{\langle \bth\rangle}
\newcommand{\thn}{{\bs \theta}_{n}}

\newcommand{\thnmu}{{\bs \theta}_{n-1}}

\newcommand{\athn}{\langle\thn\rangle}

\newcommand{\athnmu}{\langle\thnmu\rangle}

\newcommand{\Jo}{{J^\bot}}

\begin{document}

\title{Distributed Stochastic Approximation\\ for Constrained and Unconstrained Optimization\titlenote{This work is partially supported by the French National Research
Agency, under the program ANR-07 ROBO 0002}}

\author{
Pascal Bianchi\qquad  Jérémie Jakubowicz \bigskip\\
       \affaddr{Telecom ParisTech / CNRS-LTCI }\\
       \affaddr{46, rue Barrault - 75634 Paris Cedex - France}\\
       \affaddr{$\{$bianchi,jakubowi$\}$@telecom-paristech.fr}
}
\maketitle

\begin{abstract}
  In this paper, we analyze the convergence of a distributed
  Robbins-Monro algorithm for both constrained and unconstrained
  optimization in multi-agent systems.  The algorithm searches for local
  minima of a (nonconvex) objective function which is supposed to
  coincide with a sum of local utility functions of the agents.  The
  algorithm under study consists of two steps: a local stochastic
  gradient descent at each agent and a gossip step that drives the
  network of agents to a consensus.  It is proved that {\sl i)} an
  agreement is achieved between agents on the value of the estimate,
  {\sl ii)} the algorithm converges to the set of Kuhn-Tucker points
  of the optimization problem.  The proof relies on recent results
  about differential inclusions. In the context of unconstrained
  optimization, intelligible sufficient conditions are provided in
  order to ensure the stability of the algorithm.  In the latter case,
  we also provide a central limit theorem which governs the asymptotic
  fluctuations of the estimate.
  We illustrate our results in the case of distributed power allocation
  for ad-hoc wireless networks.


\end{abstract}

\section{Introduction}

The Robbins-Monro (R-M) algorithm \cite{robbins:monro:1951} is a widely used
procedure for finding the roots of an unknown function.  Its applications
range from Statistics (e.g.  \cite{delyon:lavielle:moulines:1999}),
Machine Learning (e.g.  \cite{gadat:younes:2007}),
Electrical Engineering (e.g.  \cite{widrow:mccool:larimore:johnson:1976}) 
and Communication Networks. 
Consider the problem of minimizing a given differentiable function $f$.
Formally, a R-M algorithm for that sake can be
summarized as an iterative scheme of the form $\theta_{n+1} = \theta_n +
\gamma_{n+1} (-\nabla f(\theta_n)+\xi_{n+1})$ where the sequence $(\theta_n)_{n\in\bN}$
will eventually converge to a local minimum of $f$,  and where $\xi_{n+1}$ represents a random
perturbation.

In this paper, we investigate a \emph{distributed} version of the R-M
algorithm. Distributed algorithms have aroused deep interest in the
fields of communications, signal processing, control, robotics,
computer technology, among others. The success of distributed
algorithms lies in their scalability but are often harder to analyze
than their centralized counterparts. We analyze the behavior of a
network of agents, represented as a graph, where each node/agent runs
its own local R-M algorithm and then randomly communicates with one of
its neighbors in the hope of gradually reaching a consensus over the
whole network.  One well-established device for reaching a consensus
in a network is to use gossip algorithms
\cite{bertsekas:1997}.  Since the seminal paper
of~\cite{boyd:2006}, \emph{random} gossip algorithms have been widely
studied as they encompass asynchronous networks with random switching
graph topologies.  In \cite{bertsekas:1997}, the Authors
introduce an iterative algorithm for the optimization of an objective
function in a parallel setting. The method consists in an iterative
gradient search combined with a gossip step. More recently, this
algorithm has been studied by \cite{nedic:ozdaglar:parrillo:2010,nous:icassp} in the case
where the objective function is the aggregate of some local utility
functions of the agents, assuming that a given agent is only able to
evaluate a (noisy version of) the gradient/subgradient of it own
utility function.  An alternative performance analysis is proposed by
\cite{stankovic:2008} in a linear regression perspective.

In this paper, we consider a network composed by $N\geq 1$ agents.
A given continuously differentiable utility function $f_i:\bR^d\to\bR$ 
is associated to each agent $i=1,\dots,N$, where $d$ is an integer.
We investigate the following minimization problem:
\begin{equation}
  \label{eq:pb}
  \min_{\theta\in G} \sum_{i=1}^N f_i(\theta)
\end{equation}
where $G$ is a subset of $\bR^d$ supposed to be known by each agent.
We are interested in two distinct cases: first the case of
unconstrained minimization ($G=\bR^d$), second, the case where $G$ is
a compact convex subset specified by inequality constraints. However,
we do {\bf not} suppose that the objective function $f:=\sum_i f_i$ is convex.
Moreover, we consider the context of stochastic approximation: each agent observes a
random sequence of noisy observations of the gradient $\nabla f_i$. 
We are interested in \emph{on-line} estimates of local solutions
to~(\ref{eq:pb}) using a distributed R-M algorithm.

Our contribution is the following. A distributed R-M algorithm is
introduced following~\cite{bertsekas:1997,nedic:ozdaglar:parrillo:2010,nous:icassp}.  
It is proved to converge to a consensus with probability
one (w.p.1.) that is, all agents eventually reach an agreement on their
estimate of the local solution to the minimization
problem~(\ref{eq:pb}).  In addition, each agent's estimate converges to
the set of Kuhn-Tucker points $\cL_{KT}$ of~(\ref{eq:pb}) under some
assumptions. In the unconstrained case, the proof is based on the
existence of a well-behaved Lyapunov function which ensures the
stability of the algorithm.  In the constrained case, the proof relies
on recent results of~\cite{benaim:2005} about perturbed differential
inclusions.

The paper is organized as follows. Section \ref{sec:model} introduces
the distributed algorithm and the main assumptions on the network and the observation model.
In Section \ref{sec:unconstrained}, we analyze the behavior of the algorithm in
case of unconstrained optimization ($G=\bR^d$). We prove the almost sure agreement and the
almost sure convergence of the algorithm. We provide the speed of convergence as well
as a Central Limit Theorem on the estimates. In Section \ref{sec:constrained},
we investigate the case where the domain $G$ is determined by a set of inequality constraints.
Agreement and almost sure convergence to Kuhn-Tucker points is shown.
Section \ref{sec:appli} provides an example of application to distributed power allocation for ad-hoc
wireless networks.

\section{The Distributed Algorithm}
\label{sec:model}

\subsection{Description of the Algorithm}

Each node $i$ generates a stochastic process $(\theta_{n,i})_{n\geq 1}$ in  $\bR^d$ 
using a two-step iterative algorithm:\\
\noindent {\tt [Local step]} Node $i$ generates at time
$n$ a temporary iterate $\tilde \theta_{n,i}$ given by 
\begin{equation}
  \label{eq:tempupdate}
\tilde \theta_{n,i}= P_G\left[\theta_{n-1,i} + \gamma_{n}Y_{n,i}\right]\ ,
\end{equation}
where $\gamma_n$ is a deterministic positive step size,
$Y_{n,i}$ is a random variable, and $P_G$ represents the projection
operator onto the set $G$. In particular, $P_G$ is equal to 
the identity map in case $G$ is taken to be the whole space $\bR^d$.
Random variable $Y_{n,i}$ is to be interpreted as a perturbed version of the
opposite gradient of $f_i$ at point $\theta_{n-1,i}$. As will be made clear by Assumption {\bf A}1d) below, 
it is convenient to think of
$Y_{n,i}$ as $Y_{n,i}=-\nabla f_i(\theta_{n-1,i})+\delta M_{n,i}$ where $(\delta M_{n,i})_n$
is a martingale increment sequence which stands for the random perturbation.

\noindent {\tt [Gossip step]} Node $i$ is 
able to observe the values $\tilde\theta_{n,j}$ of some other $j$'s 
and computes the weighted average:
$$
\theta_{n,i}=\sum_{j=1}^N w_n(i,j)\,\tilde \theta_{n,j}
$$
where $W_n := [ w_n(i,j) ]_{i,j=1}^N$ is a stochastic matrix.

We cast this algorithm into a more compact vector form.
Define the random vectors $\thn$ and
$Y_n$ as $\thn:=(\theta_{n,1}^T,\dots,\theta_{n,N}^T)^T$ and $Y_n =
(Y_{n,1},\dots, Y_{n,N})^T$  where $^T$ denotes transposition.  The algorithm reduces to:\\
\framebox{\parbox{\columnwidth}{\begin{equation}
\thn = (W_n\otimes I_d)P_{G^{N}}\left[\thnmu+\gamma_{n}Y_n\right]
\label{eq:algo} 
\end{equation}}}\\
where $\otimes$ denotes the Kronecker product, $I_d$ is the $d× d$ identity matrix
and $P_{G^{N}}$ is the projector onto the $N$th order product set $G^N:=G× \cdots× G$.


\subsection{Observation and Network Models}
\label{sec:assumptions}


The time-varying communication network between the nodes is represented by
the sequence of random matrices $(W_n)_{n\geq 1}$. 
For any $n\geq 1$, we introduce the $\sigma$-field $\cF_n = \sigma({\bs \theta}_0,Y_{1:n},W_{1:n})$.
The distribution of the random vector $Y_{n+1}$ conditionally to $\cF_n$ is assumed to be such that:
$$
\bP\left( Y_{n+1} \in A \, |\, \cF_n\right) = \mu_{\thn}(A)
$$
for any measurable set $A$, where $\left(\mu_\bth\right)_{\bth\in \bR^{dN}}$ is a given family
of probability measures on $\bR^{dN}$. For any $\bth\in\bR^{dN}$, define $\bE_{\bth}[g(Y)]:= \int g(y)\mu_{\bth} (dy)$.
Denote by $\un$ the $N× 1$ vector whose components are all equal to one.
Denote by $|x|$ the Euclidean norm of any vector $x$. It is assumed that:
\begin{assum}
\label{hyp:model} The following conditions hold:\\
 a) Matrix $W_n$ is doubly stochastic: $W_n\un=W_n^T\un=\un$.\\
 b) $(W_n)_{n\geq 1}$ is a sequence of square-integrable matrix-valued random variables.
The spectral radius $\rho_n$ of matrix $\bE(W_nW_n^T)-\un\un^T/N$ satisfies: 
\begin{equation*}
\lim_{n\to\infty}n(1-\rho_n)=+\infty\ .
\label{eq:limrhon}
\end{equation*}
 c) For any positive measurable functions $g_1,g_2$, 
\begin{equation*} 
  \bE[g_1(W_{n+1})g_2(Y_{n+1})\vert \cF_n] 
=\bE[g_1(W_{n+1})]\, \bE_{\thn}[g_2(Y)]\ .
\end{equation*}
 d) $\bth_0\in G^N$ and $\bE[|{\bs\theta}_0|^2]<+\infty$.\\
 e) For any $i=1,\dots, N$, $f_i$ is continuously differentiable.\\
 f) For any $\bth=(\theta_1^T,\cdots,\theta_N^T)^T$, 
$$
\bE_\bth[Y] = -(\nabla f_1(\theta_1)^T,\cdots,\nabla f_N(\theta_N)^T)^T\ .
$$
\end{assum}

Condition {\bf A\ref{hyp:model}}a) is satisfied provided that the
nodes coordinate their weights. Coordination schemes are discussed
in~\cite{nedic:ozdaglar:parrillo:2010,boyd:2006}. 
Due to {\bf A\ref{hyp:model}}b), note that $\rho_n<1$ as soon as $n$
is large enough. Loosely speaking, Assumption {\bf A\ref{hyp:model}}b)
ensures  that $\bE(W_nW_n^T)$ is close enough to the projector $\un\un^T/N$ on the line $\{t\un : t\in\bR\}$.
This way, the amount of information exchanged in the network remains sufficient
in order to reach a consensus. 
Condition {\bf A\ref{hyp:model}}c) implies that r.v. $W_{n+1}$ and $Y_{n+1}$
are independent conditionally to the past. 
In addition, $(W_n)_{n\geq 1}$ forms an independent sequence (not necessarily identically distributed).
Condition {\bf A\ref{hyp:model}}f) means that each $Y_{n,i}$ can be interpreted as
a `noisy' version of $-\nabla f_i(\theta_{n-1,i})$. The distribution of the random additive perturbation $Y_{n,i}+\nabla f_i(\theta_{n-1,i})$
is likely to depend on the past through the value of $\thnmu$, but has a zero mean for any given 
value of $\thnmu$.
\begin{assum} \label{hyp:step} 
 a) The deterministic sequence $(\gamma_n)_{n\geq 1}$ is positive and such that $\sum_n\gamma_n=\infty$.\\
 b) There exists $\alpha>1/2$ such that:
\begin{align}
&\lim_{n\to\infty} n^{\alpha}\gamma_n = 0 \label{eq:condgamma}\\
&\liminf_{n\to\infty} \frac{1-\rho_n}{n^{\alpha}\gamma_n}>0 \ . \label{eq:condrho}
\end{align}
\end{assum}
Note that, when~(\ref{eq:condgamma}) holds true then
$\sum_n\gamma_n^2 <\infty$, which is a rather common assumption in the framework
of decreasing step size stochastic algorithms~\cite{kushner:2003}. 
In order to have some insights on~(\ref{eq:condrho}), consider the case
where $1-\rho_n=a/n^\eta$ and $\gamma_n=\gamma_0/n^\xi$ for some constants $a,\gamma_0>0$.
Then, a sufficient condition for~(\ref{eq:condrho}) and~{\bf A\ref{hyp:step}}a) is:
$$
0\leq \eta < \xi -1/2 \leq 1/2\ .
$$
In particular, $\xi\in (1/2,1]$ and $\eta\in [0,1/2)$. The case $\eta=0$ typically
correspond to the case where matrices $W_n$ are identically
distributed.  In this case, $\rho_n=\rho$ is a constant w.r.t. $n$ and our
assumptions reduce to: $\rho <1$. However, matrices $W_n$ are not
necessarily supposed to be identically distributed. Our results hold
in a more general setting. As a matter of fact, all results of this
paper hold true when matrices $W_n$ are allowed to converge to the
identity matrix (but at a moderate speed, slower than $1/\sqrt{n}$ in
any case). Therefore, matrix $W_n$ may be taken to be the identity
matrix with high probability, without any restriction on the results
presented in this paper. From a communication network point of view,
this means that the exchange of information between agents reduces to
zero as $n\to\infty$. This remark has practical consequences in case of
wireless networks, where it is often required to limit as much as
possible the communication overhead.


\section{Unconstrained Optimization}
\label{sec:unconstrained}

\subsection{Framework and Assumptions}
In this section, $G$ is taken to be the whole space, so that the algorithm~(\ref{eq:algo})
simplifies to:
\begin{equation}
\thn = (W_n\otimes I_d)\left(\thnmu+\gamma_{n}Y_n\right)\ .
\label{eq:unconstrained}
\end{equation}
Our aim is to study the convergence of the above iterate sequence.
Note that sequence $\thn$ is not {\sl a priori} supposed to stay in a compact set.
Additionally, in most  situations, large values of some components of $\thnmu$
may lead to large values of $Y_n$. Otherwise stated, 
one of the main issues in the unconstrained case is to demonstrate 
the {\bf stability} of the algorithm~(\ref{eq:unconstrained})
based on explicit and intelligible assumptions on the objective function $f$ and on the stochastic perturbation.

\begin{assum} \label{hyp:V} 
There exists a  function $V : \bR^d \to \bR^+$ such that:\\
 a) $V$ is differentiable and  $\nabla V$ is a Lipschitz function.\\
 b) For any $\theta \in \bR^d$, $ -\nabla V(\theta)^T \nabla f(\theta) \leq 0$.\\ 
 c) There exists a constant $C_1$, such that for any $\theta \in \bR^d$, $|\nabla V(\theta)|^2\leq C_1(1+V(\theta))$.\\
 d) For any $M>0$, the level sets $\{\theta\in\bR^d: V(\theta)\leq M \}$ are compact. \\
 e) The set $\cL:= \{\theta\in \bR^d: \nabla\!V(\theta)^T \nabla f(\theta) =0 \}$ is bounded.\\ 
 f) $V(\cL)$ has an empty interior.
\end{assum}
Assumption {\bf A\ref{hyp:V}}b) means that $V$ is a Lyapunov function for $-\nabla f$. 
In case of gradient systems obtained from optimization problems such as~(\ref{eq:pb}), a Lyapunov function $V$
is usually given by the objective function $f$ itself, or by a composition $\phi \circ f$ of $f$ with
a well-chosen increasing map $\phi$: Assumption {\bf A\ref{hyp:V}}b) is then trivially satisfied. In this case,
the set $\cL$ reduces to the roots of $\nabla f$:
$$
\cL = \{\theta\in \bR^d: \nabla f(\theta) =0 \}\ .
$$
Assumption {\bf A\ref{hyp:V}} combined with the condition $\sum_n \gamma_n = +\infty$
  allows to prove the convergence of the deterministic sequence  $t_{n+1} = t_n
  - \gamma_{n+1} \nabla f(t_n)$ to the set $\mathcal{L}$. When $\nabla f$ is unknown and
  replaced by a stochastic approximation, the limiting behavior of the
  noisy algorithm is similar under some regularity conditions
  and under the assumption that the step-size sequence satisfies $\sum_n \gamma_n^2 < \infty$.
 Assumption {\bf A\ref{hyp:V}}c) implies that $V$
increases at most at quadratic rate $O(|\theta|^2)$ when $|\theta|\to\infty$.
Assumption~{\bf A\ref{hyp:V}}f) is trivially satisfied when $\cL$ is finite.

We denote by $J:=(\un\un^T/N)\otimes I_d$ the projector onto the consensus subspace
$\left\{ \un\otimes \theta : \theta\in\bR^d\right\}$
and by $J^\bot:=I_{dN}-J$ the projector onto the orthogonal subspace. 
For any vector $\bth\in\bR^{dN}$, remark that
$\bth =\un\otimes \ath+J^\bot\bth$ where
\begin{equation}
  \label{eq:average}
  \ath := \frac 1N ({\un^T\otimes I_d})\bth 
\end{equation}
is a vector of $\bR^d$ equal to $(\theta_1+\dots+\theta_N)/N$ in case we
write $\bth=(\theta_1^T,\dots,\theta_N^T)^T$ for some $\theta_1,\dots, \theta_N$ in $\bR^d$.

\begin{table}[t]
  \centering
  \begin{tabular}[t]{|c|l|}
    \hline 
    $\theta$ & dummy variable in $\bR^d$ \\
    $\bth$ &  dummy variable in $\bR^{dN}$ \\
    $\theta_{n,i}$ & estimate at agent $i$ and at time $n$ in $\bR^d$ \\
    $\thn$ & vector of the $N$ agents estimates in $\bR^{dN}$ \\
    $\athn$ & average of the agents estimates in $\bR^{d}$ \\
    $J$ & projector onto the consensus subspace \\
    $\Jo\thn$ & disagreement vector between agents in $\bR^{dN}$ \\
    $f$ & Aggregate utility function $f=\sum_i f_i$ \\
    $Y_n$ & vector of all observations at time $n$, in $\bR^{dN}$ \\
    $\un$ & Vector $(1,\cdots,1)^T$ in $\bR^N$\\
    $\gamma_n$ & step size \\
    $\rho_n$ & spectral radius of $\bE(W_nW_n^T)-\un\un^T/N$\\ 
    \hline
  \end{tabular}
  \caption{Summary of useful notations}
\end{table}

\begin{assum} \label{hyp:H} 
  {\sl There exists a constant $C_2$, such that for
    any $\bs\theta=(\theta_1^T,\cdots,\theta_N^T)^T$ in $\bR^{dN}$, 
\begin{align}
&\bE_{\bth}\left[\,\left|Y\right|^2 \right] \leq C_2\left(1 + V(\ath) + |J^\bot\bth|^2 \right) \label{eq:Hquadra} \\
& \left| \nabla f(\ath)-\frac 1N\sum_{i=1}^N \nabla f_i(\theta_i) \right| \leq C_2 |J^\bot\bth|\label{eq:JHLipsch} 
\end{align}
}
\end{assum}
Condition~(\ref{eq:Hquadra}) implies
that $|\nabla f(\theta)|^2\leq C_2(1 + V(\theta))$. 
This means that the mean field $\nabla f(\theta)$
cannot increase more rapidly than $O(|\theta|)$ as $|\theta|\to\infty$.
Condition~(\ref{eq:JHLipsch}) is in particular satisfied in case all $f_i$'s are Lipschitz function.
Condition~(\ref{eq:JHLipsch})  ensures that small variations of vector $\bth$
near the consensus space cannot produce large variations of~$\sum_i\nabla f_i(\theta_i)$.


\subsection{Convergence w.p.1}
\label{sec:consensus}

The disagreement between agents can be quantified through the norm of the vector
$$
\Jo\thn = \thn-\un\otimes\athn\ .
$$
\begin{lemma}[Agreement]
\label{lem:agreement}
  Under {\bf A\ref{hyp:model}-\ref{hyp:step}}, {\bf A\ref{hyp:V}}a-c) and {\bf A\ref{hyp:H}},
{\sl i)} $\Jo\thn$ converges to zero almost surely (a.s.) as $n\to\infty$.\\
{\sl ii)} For any $\beta<2\alpha$, $\lim_{n\to\infty} n^\beta \bE\left[|\Jo\thn|^2\right]=0\ .$
\end{lemma}
Lemma~\ref{lem:agreement} is the key result to characterize the asymptotic behavior
of the algorithm. The proof is omitted due to lack of space, but will be presented
in an extended version of this paper.
Point {\sl i)} means that the disagremeent between agents converges 
almost-surely to zero. Point {\sl ii)} states that the convergence also holds in $L^2$
and that the convergence speed is faster than $1/\sqrt{n}$: 
This point will be revealed especially useful in Section~\ref{sec:clt}.
Define $\sd(\theta,A):=\inf\{ |\theta-\varphi|\,:\varphi \in A\}$ for any $\theta\in\bR^d$ and $A\subset\bR^d$.
Define $\un\otimes\cL:=\{\un\otimes\theta\ :\ \theta\in\cL\}$.
\begin{theo}
\label{the:cv}
Assume {\bf A\ref{hyp:model}}, {\bf A\ref{hyp:step}}, {\bf A\ref{hyp:V}} and
{\bf A\ref{hyp:H}}. Then, w.p.$1$,
\[
\lim_{n\to\infty}\sd(\thn,\un\otimes\cL)=0 \ .
\]
Moreover, w.p.$1$, $(\athn)_{n\geq 1}$ converges to a connected component of
$\cL$.
\end{theo}
Theorem~\ref{the:cv} states that, almost surely, the vector of iterates
$\thn$ converges to the consensus space as $n\to\infty$. Moreover, the average
iterate $\athn$ of the network converges to some connected component of $\cL$.
When $\cL$ is finite, Theorem~\ref{the:cv} implies that $\thn$ converges a.s. to some point in~$\un\otimes\cL$.

The proof of Theorem~\ref{the:cv} is omitted. Conditions {\bf
  A\ref{hyp:step}}, {\bf A\ref{hyp:V}}a-e) and {\bf A\ref{hyp:H}} imply that, almost-surely, \textit{(a)} the sequence
$(\athn)_{n \geq 1}$ remains in a neighborhood of
$\mathcal{L}$ thus implying that the sequence remains in a compact set
of $\bR^d$ and \textit{(b)} the sequence $(V(\athn))_{n \geq 1}$
converges to a connected component of
$V(\mathcal{L})$.  Finally, {\bf A\ref{hyp:V}}f) implies the
convergence of $(\athn)_{n \geq 1}$ to a connected component of~$\mathcal{L}$.

\subsection{Central Limit Theorem}
\label{sec:clt}

Let $\theta_*$ be a point satisfying the following Assumption.
\begin{assum} \label{hyp:clt} 
a) $\theta_*\in\cL$.\\
b) Function $f$ is two times differentiable at point $\theta_*$ and
$f(\theta)=H(\theta_*)(\theta-\theta_*)+ O( |\theta-\theta_*|^2 )$
for any $\theta$ in a neighborhood of $\theta_*$, where 
$H(\theta_*)$ denotes the $d× d$ Hessian matrix of $f$ at point~$\theta_*$.\\
c) $H(\theta_*)$ is a stable matrix: the largest real part of its eigenvalues
is $-L$, where $L>0$.\\
d) There exists $\delta>0$ such that the function:
$\bth\mapsto \bE_{\bth}\left[\left|Y\right|^{2+\delta} \right]$ is bounded in a neighborood of $\un\otimes \theta_*$. \\ 
e) The matrix-valued function  $Q:\bR^{dN} \to \bR^{d× d}$ defined by:
\begin{equation*}
Q(\bth)=\bE_\bth\big[\left(\la Y\ra - \bE_\bth\la Y\ra\right)\left(\la Y\ra - \bE_\bth\la Y\ra\right)^T\big]
\end{equation*}
is continuous at point $\un\otimes\theta_*$.\\ 
f) Matrix $Q(\un\otimes\theta_*)$ is positive definite.
\end{assum}

\begin{assum}
\label{hyp:stepCLT}
For any $n\geq 1$, $\gamma_n = \gamma_0\, n^{-\xi}$ where $\xi\in (1/2,1]$ and $\gamma_0>0$.
In case $\xi=1$, we furthermore assume that $2L\gamma_0>1$.
\end{assum}

The normalized disagreement vector 
$\gamma_n^{-1/2}\Jo\thn$ converges to zero in probability by Lemma~\ref{lem:agreement}ii).
Therefore, it can be shown that the asymptotic analysis reduces to the study of the average
$\athn$. To that end, we remark from  {\bf A\ref{hyp:model}}a) that $(\un\otimes I_d)(W_n\otimes I_d)=(\un\otimes I_d)$.
Thus, $\athn$ satisfies:
$\athn = \athnmu + \gamma_{n} \la Y_n\ra$.
The main step is to rewrite
the above equality under the form:
$$
  \athn = \athnmu+\gamma_n \left(-\nabla f(\athnmu)+\delta\tilde M_n+r_n\right)\ ,
$$
where $\delta\tilde M_n$ is a martingale increment sequence satisfying some desired
properties (details are omitted) and where $r_n$ is a random sequence
which is proved to be negligible.  The final result is a consequence
of~\cite{pelletier:1998}. A sequence of r.v.  $(X_n)_n$ is
said to converge in distribution (stably) to a r.v. $X$ given an event
$E$ whenever $ \lim_n\bE\left( g(X_n)\un_E\right) = \bE\left(
  g(X)\right)\bP(E) $ for any bounded continuous function $g$.
\begin{theo}
  Assume {\bf A\ref{hyp:model}--\ref{hyp:H}}, {\bf A\ref{hyp:stepCLT}} and assume that there
  exists a point $\theta_*$ satisfying {\bf A\ref{hyp:clt}}.
Then, given the event
$$
\{\lim_{n\to\infty} \la\thn\ra = \theta_*\}\ ,
$$
the following holds true:
$$
\gamma_n^{-1/2} \left(\thn-\un\otimes \theta_*\right) \xrightarrow[]{\cD} \un \otimes Z\ .
$$
where $Z$ is a $d× 1$ zero mean Gaussian vector whose covariance matrix $\Sigma$
is the unique solution~to:
\begin{equation}
  \label{eq:lyapSigma}
  \left(H(\theta_*)+\zeta I_d\right)\Sigma + \Sigma   \left(H(\theta_*)+\zeta I_d\right) = -Q(\un\otimes\theta_*)
\end{equation}
where $\zeta=0$ if $\xi \in (1/2,1)$ and $\zeta=1/(2\gamma_0)$ if $\xi=1$.
\label{the:clt}\end{theo}

Theorem~\ref{the:clt} states that, given the event that sequence $\thn$
converges to a given point $\un\otimes\theta_*$, the normalized error
$\gamma_n^{-1/2}(\thn-\un\otimes\theta_*)$ converges to a Gaussian vector.
The latter limiting random vector belongs to the consensus subspace {\sl i.e.},
it has the form $\un\otimes Z$, where $Z$ is a Gaussian r.v. of dimension $d$.
Theorem~\ref{the:clt} has the following important consequences.
First, thanks to the gossip step, the component of the error vector
in the orthogonal consensus subspace is asymptotically negligible.
The dominant source of error is due to the presence of observation noise in
the algorithm, and not on possible disagreements between agents.
As a matter of fact, the limiting behavior of the average estimate
is similar to the one that would have been observed in a centralized setting.
Interestingly, this remark is true even if the agents reduce their cooperation
as time increases (consider the case where $W_n=I_d$ with probability converging to one).

\subsection{Influence of the network topology}
\label{subsec:network}

To illustrate our claims, assume for simplicity that $(W_n)_{n\geq 1}$ is an i.i.d. sequence.
Then $\rho_n=:\rho$ is a constant w.r.t. $n$. In this case, all our hypotheses on sequence $(W_n)_{n\geq 1}$
reduce~to:
\begin{equation}
\label{eq:rhoinfun}
\rho < 1\ .
\end{equation}
In order to have more insights, it is useful to relate the above inequality to a connectivity
condition on the network. To that end, we focus on an example.
Assume for instance that matrices $W_n$ follow the now widespread
asynchronous random pairwaise gossip model described in~\cite{boyd:2006}. 
At a given time instant $n$,
a node $i$, picked at random, wakes up and exchange information with an other node $j$ also
chosen at random (other nodes $k\notin \{i,j\}$ do not participate to any exchange of information).
$W_n$ belongs to the alphabet $\{\sW_{i,j} : i,j=1,\dots,N\}$ where:
$$
\sW_{i,j} := I_d - (e_i-e_j)(e_i-e_j)^T/2\ ,
$$
where $e_i$ represents the $i$th vector of the canonical basis
($e_i(k)=1$ if $i=k$, zero otherwise).  Denote by $P_{i,j}=P_{j,i}$
the probability that the active pair of nodes at instant $n$ coincides
with the pair $\{i,j\}$.  In practice, $P_{i,j}$ is nonzero only if
nodes $i,j$ are able to communicate ({\sl i.e.} they are connected).
Consider the weighted nondirected graph $\cG=(\cE,\cV,\cW)$ where $\cE$ is the set of
vertices $\{1,\dots,N\}$, $\cV$ is
the set of edges (by definition, $i$ is connected to $j$ iff
$P_{i,j}>0$), and $\cW$ associates the weight $P_{i,j}$ to the
connected pair $\{i,j\}$.  Using \cite{boyd:2006}, it is straightforward to show that
condition~(\ref{eq:rhoinfun}) is equivalent to the condition
that $\cG$ is connected.
\begin{coro}
  Replace conditions~(\ref{eq:limrhon}) and~(\ref{eq:condrho}) with the
assumption that $\cG$ is connected. Then Theorems~\ref{the:cv} and~\ref{the:clt}
still hold true.
\end{coro}
In particular, the (nonzero) spectral gap of the Laplacian of $\cG$ has no impact on the
asymptotic behavior of sequence $\thn$. 
Stated differently, the dominant source of error in the asymptotic regime is due 
to the observation noise. The disagreement between agents is negligible even in
networks with a low level of connectivity.

\section{Constrained Optimization}
\label{sec:constrained}
\subsection{Framework and Assumptions}
We now study the case where the set $G$ is
determined by a set of $p$ inequality constraints ($p\geq 1$):
\begin{equation}
\label{eq:G}
G := \left\{ \theta\in\bR^d\ :\ \forall j=1,\dots,p,\ q_j(\theta)\leq 0\right\}
\end{equation}
for some functions $q_1,\dots,q_p$ which satisfy the following conditions.
Denote by $\dG$ the boundary of $G$. For any $\theta\in G$, we denote by $A(\theta)\subset \{1,\dots,p\}$
the set of active constraints \emph{i.e.}, $q_j(\theta)=0$ if $j\in A(\theta)$ and $q_j(\theta)<0$ otherwise.
\begin{assum}
\label{hyp:KT}
a) The set $G$ defined by~(\ref{eq:G}) is compact.\\
b) For any $j=1,\dots,p$,  $q_j:\bR^d\to\bR$ is a convex function\\
c) For any $j=1,\dots,p$,  $q_j$ is two times
continuously differentiable in a neigborhood of $\dG$.\\
c) For any $\theta\in\dG$,  $\{\nabla q_j(\theta):j\in A(\theta)\}$ is a linearly independent collection of vectors.\\
\end{assum}
In the particular case where all utility functions $f_1,\dots,f_N$ are assumed convex,
it is possible to study the convergence w.p.1 of the algorithm~(\ref{eq:algo})
following an approach similar to~\cite{nedic:ozdaglar:parrillo:2010}, and to prove under some conditions that consensus is achieved
at a global minimum of the aggregate objective function $f$. Nevertheless, 
utility functions may not be convex in a large number of situations, and there seems to be few hope
to generalize the proof of~\cite{nedic:ozdaglar:parrillo:2010} in such a wider setting. In this paper, we do {\bf not} assume
that the utility functions are convex. In this situation, convergence to a global minimum of~(\ref{eq:pb})
is no longer guaranteed. We nevertheless prove the convergence of the algorithm~(\ref{eq:algo})
to the set of Kuhn-Tucker (KT) points $\cL_{KT}$:
$$
\cL_{KT} := \left\{ \theta \in G\ : \ -\nabla f(\theta) \in \cN_G(\theta)\right\}\ ,
$$
where $\cN_G(\theta)$ is the normal cone to $G$ \emph{i.e.},
$\cN_G(\theta):=\{ v \in \bR^d\ :\ \forall \theta'\in G, v^T (\theta-\theta')\geq 0\}$.
To prove convergence, we need one more hypothesis: 
\begin{assum}
  \label{hyp:regc} The following two conditions hold:\\
a) $\sup_{\bth\in G^N}\bE_\bth[\,|Y|^2]<\infty$.\\
b) Inequality~(\ref{eq:JHLipsch}) holds for any $\bth\in G^N$.
\end{assum}

\subsection{Convergence w.p.1}

Theorem~\ref{the:cvc} below establishes two points: First, a consensus
is achieved as $n$ tends to infinity, meaning that $\Jo\thn$ converges a.s. to zero.
Second, the average estimate $\athn$ converges to the set of KT points.
\begin{theo}
\label{the:cvc}
Assume {\bf A\ref{hyp:model}}, {\bf A\ref{hyp:step}},
{\bf A\ref{hyp:KT}} and {\bf A\ref{hyp:regc}}. 
Then, w.p.$1$,
\[
\lim_{n\to\infty}\sd(\thn,\un\otimes\cL_{KT})=0 \ .
\]
Moreover, w.p.$1$, $(\athn)_{n\geq 1}$ converges to a connected component of
$\cL_{KT}$.
\end{theo}

As a consequence, if $\cL_{KT}$ contains only isolated points, sequence 
$\athn$ converges almost surely to one of these points.
The complete proof of Theorem~\ref{the:cvc} is omitted.
We however provide some elements of the proof in the next paragraph.

\subsection{Sketch of the proof}

To simplify the presentation,
we shall focus on the case $p=1$ \emph{i.e.}, there is only one inequality constraint. We put $q:=q_1$
and define $e:=\nabla q/|\nabla q|$ the normalized gradient of function $q$ ($e$ is well defined
in a neighborhood of $\dG$ by {\bf A\ref{hyp:KT}}c)).

{\bf Step 1:} {\it Agreement is achieved as $n\to\infty$.}\\
Similarly to the unconstrained optimization case (recall previous Lemma~\ref{lem:agreement}), 
the first step of the proof of Theorem~\ref{the:cvc}
is to establish that $|\Jo\thn|$ converges a.s. to zero. As a noticeable difference with
the unconstrained case, here stability issues do not come into play as $G$ is bounded
(for this reason, the proof of agreement is simpler than in the unconstrained case).

{\bf Step 2:} {\it Expression of the average $\athn$ in a R-M like form}.\\
Using $(\un\otimes I_d)(W_n\otimes I_d)=(\un\otimes I_d)$, it is convenient to write
$\athn = \athnmu +\gamma_n Z_n$
where 
$$
Z_n := \frac{1}{\gamma_nN}\sum_{i=1}^NP_G(\theta_{n-1,i}+\gamma_n Y_{n,i})-\theta_{n-1,i}\ .
$$
Consider the martingale increment sequence $\Delta_n:=Z_n-\bE(Z_n|\cF_{n-1})$.
From Assumption {\bf A\ref{hyp:regc}}a), it can be shown that $\sup_n \bE[\,|\Delta_n|^2]<\infty$.
Now note that for any $\theta\in G$, $y\in\bR^d$,
\begin{equation}
\lim_{\gamma\downarrow0} \gamma^{-1}\left(P_G(\theta+\gamma y)-\theta\right)=y-(y^Te(\theta))^+e(\theta){\bs 1}_{\dG}(\theta)\ ,
\label{eq:limP}
\end{equation}where $(x)^+:=\max(x,0)$ and where ${\bs 1}_{\dG}$ is the indicator function of $\dG$.
Using~(\ref{eq:limP}) along with {\bf A\ref{hyp:KT}}c) and {\bf A\ref{hyp:regc}}b)
and the fact that $|\Jo\thn|$ converges to zero, we obtain after some algebra:
\begin{equation}
\athn = \athnmu + \gamma_n h(\thnmu) +\gamma_n\Delta_n+\gamma_n u_n
\label{eq:rmc}
\end{equation}
where $u_n$ is some sequence which converges to zero a.s. and where we defined for any $\bth\in G^N$,
$$
h(\bth):=-\nabla f(\ath)- \frac{e(\ath)}{N}\sum_{i=1}^N\bE_{\bth}\left[(Y_{i}^Te(\theta_{i}))^+\right]{\bs 1}_{\dG}(\theta_{i})\ .
$$

{\bf Step 3:} {\it From equality to inclusion}.\\
Equality~(\ref{eq:rmc}) is still far from a conventional R-M equation.
Indeed, the second term of the righthand side $\gamma_n h(\thnmu)$ is not a function of the average
$\athnmu$ as it depends on the whole vector $\thn$. Of course, since the agreement is achieved for large $n$,
$\thnmu$ should be close to $\un\otimes \athnmu$. If $h$ were continuous, one could thus write
$h(\thnmu)\simeq h(\un\otimes\athnmu)$ solving this way the latter issue. This is unfortunately not the case,
due to the presence of indicator functions in the definition of $h$. We must resort to inclusions.
For any $\epsilon\geq 0$ and any $\theta\in G$, define the following subset of $\bR^d$:
$$
F_\epsilon (\theta) := \left\{ -\nabla f(\theta) - x\, e(\theta) {\bs 1}_{\sd(\theta,\dG)\leq \epsilon}\ :\ x\in [0,M]\right\}
$$
where $M<\infty$ is a fixed constant chosen as large as needed, and where
${\bs 1}_{\sd(\theta,\dG)\leq \epsilon}$ is equal to one if $\theta$ is at distance less than $\epsilon$ of the boundary,
and to zero otherwise. In particular, ${\bs 1}_{\sd(\theta,\dG)\leq \epsilon}={\bs 1}_{\dG}(\theta)$ for $\epsilon=0$.
It is straightforward to show that:
$$
\forall \bth\in G^N,\ h(\bth)\in F_{|\Jo\bth|}(\ath)
$$
provided that $M$ is chosen large enough. Finally, equality~(\ref{eq:rmc}) can be interpreted
in terms of the following inclusion:
\begin{equation}
  \label{eq:incl}
  \athn \in \athnmu+\gamma_n F_{\epsilon_n}(\athnmu)+\gamma_n\Delta_n+\gamma_nu_n
\end{equation}
where we defined for simplicity $\epsilon_n:=|\Jo\thnmu|$.

{\bf Step 4:} {\it Interpolated process and differential inclusions.}\\
From this point to the end of the proof, we shall now study one fixed trajectory 
$(\la\thn(\omega)\ra)_n$ of the random process $\la\thn\ra$, where $\omega$ belongs to an event
of probability one such that $\epsilon_n(\omega)\to 0$, $u_n(\omega)\to 0$ as $n$ tends to infinity, 
and sequence $(\Delta_n(\omega))_n$ satisfies some asymptotic
rate of change condition (see~\cite{kushner:2003,benaim:2005} for details).
Dependencies in $\omega$ are however omitted for simplicity.
Motivated by the approach of~\cite{benaim:2005}, we consider the following 
continuous-time interpolated process. Define $\tau_n=\sum_{k=1}^n\gamma_k$ and
$$
\Theta(t) := \athnmu + \frac{\athn-\athnmu}{\tau_n-\tau_{n-1}}(t-\tau_n)\,,\quad \tau_{n-1}\leq t <\tau_n\ .
$$
The next step is to prove that $\Theta$ is a \emph{perturbed solution} 
 to the differential inclusion:
 \begin{equation}
\frac{dx(t)}{dt} \in F_0(x(t))\ .\label{eq:di}
\end{equation}
When we write that $x$ is a solution to~(\ref{eq:di}), 
we mean that $x$ is an absolutely continuous mapping $x:\bR\to\bR^d$
such that~(\ref{eq:di}) is satisfied for almost all $t\in\bR$.
A function $\Theta$ is a perturbed solution to~(\ref{eq:di}) if it 
`shadows' the behavior of a solution to~(\ref{eq:di})
as $t\to\infty$ in a sense made clear in~\cite{benaim:2005}. In order to prove
that $\Theta$ is a perturbed solution to~(\ref{eq:di}), the materials
are close to those of~\cite{benaim:2005} (see Proposition~1.3) with some care, however,
about the fact that the mean field $F_{\epsilon_n}$ is nonhomogeneous in our context (it depends
on time $n$).
The proof is concluded by straightforward application of~\cite{benaim:2005}.
Consider the differential inclusion~(\ref{eq:di}):
function $f$ is a Lyapunov function for the set of KT points $\cL_{KT}$.
Therefore, by~\cite{benaim:2005}, the limit set 
$$
\bigcap_{t\geq 0} \overline{\Theta\left([t,+\infty)\right)}
$$
is included in $\cL_{KT}$. This concludes the proof.
\bigskip
\bigskip

\section{Application: Power Allocation}
\label{sec:appli}

\subsection{Framework}

The context of power allocation for wireless networks has recently raised a great deal of attention
in the field of distributed optimization,
cooperative and noncooperative game theory (see~\cite{samson:2011} and references therein).
We consider an \emph{ad hoc} network composed of $N$ transmit-destination pairs.  
Each agent/user sends digital data to its receiver through $K$ parallel (sub)channels. 
The channel gain of the $i$th user
at the $k$th subchannel is represented by a positive coefficient
$A^{i,i;k}$ which can be
interpreted as the square modulus of the corresponding complex valued
channel gain.  As all agents share the same spectral band, user $i$
suffers from the multiuser interference produced by other users $j\neq
i$. Denote by $p^{i;k}\geq 0$ the power allocated by user $i$ to the $q$th subchannel.
We assume that $\sum_{k=1}^K p^{i;k}\leq \cP_i$ where $\cP_i$ is the maximum allowed power for user $i$.
Define $p^i=[p^{i;1},\cdots,p^{i;K}]^T$ and $\theta = [{p^1}^T,\cdots,{p^N}^T]^T$ the vector of all powers of all users
of size $d:=KN$. 
Assuming deterministic channels, user $i$ is able to provide its
destination with rate given by (see \emph{e.g.}
\cite{Scutari-Palomar-09})
$$
R_i(\theta, A^i) := \sum_{k=1}^K\log\left(1+\frac{A^{i,i;k}p^{i;k}}{\sigma_i^2+\sum_{j\neq i}A^{j,i:k}p^{j;k}}\right)
$$
where $A^{j,i:k}$ is the (positive) channel gain between transmitter
$j$ and the destination of the $i$th transmit-destination pair, and
where $A^i=[A^{1,i;1},\cdots,A^{N,i:K}]^T$.  Here, $\sigma_i^2$ is the variance
of the additive white Gaussian noise at the destination of source $i$.
The aim is to select a relevant value for the resource allocation
parameter $\theta\in G$ in a distributed fashion, where $G$ is the set of
constraints obtained from the aforementioned power constraints
$\cP_1,\dots,\cP_N$ and positivity constraints.

\subsection{Deterministic Coalitional Allocation}

To simplify the presentation, we first consider the case of fixed deterministic channel gains $A^1,\dots,A^N$.
A widespread approach consists in computing $\theta$ through
the so-called best response dynamics. At every step of the
iteration, an agent $i$ updates its own power vector $p^{i}$ assuming
other users' power to be fixed. This is the
well known iterative water filling algorithm~\cite{Scutari-Palomar-09}.
Here, we are interested in a different perspective. The aim is rather to search for
social fairness between users. We aim at 
finding a local maximum of the following weighted sum rate:
\begin{equation}
\sum_{i=1}^N \beta_i\, R_i(\theta,A^i)
\label{eq:sumrate}
\end{equation}
where $\beta_i$ is an arbitrary positive deterministic weight known only by agent $i$.
Consider the following deterministic gradient algorithm.
Each user $i$ has an estimate $\theta_{n,i}$ of $\theta$ at the $n$th iteration.
Here, we stress the fact that a given user has not only an estimate of what should
be its own power allocation $p^i$, but has also an estimate of what should be the power allocation
of other users $j\neq i$. Denote by $\thn=[\theta_{n,1}^T,\cdots,\theta_{n,N}^T]^T$ the vector of size $dN=KN^2$
which gathers all local estimates. Similarly to~(\ref{eq:algo}), a distributed algorithm
for the maximization of~(\ref{eq:sumrate}) would have the form
$\thn = (W_n\otimes I_d)P_{G^{N}}\left[\thnmu + \gamma\,Y(\thnmu;A) \right]$
where $Y(\bth;A)=[\beta_1 \nabla_\theta R_1(\theta_1;A^1)^T,\cdots,\beta_N \nabla_\theta R_N(\theta_N;A^N)^T]^T$ and 
where $\nabla_\theta$ is the gradient operator with respect to the first argument $\theta$ of $R_i(\theta,A^i)$.

\subsection{Stochastic Coalitional Allocation}
In many situations however, the above algorithm is impractical.
This is for instance the case when the channel gains are random and rapidly time-varying
in an ergodic fashion. This is also the case when channel gains are known only up to a random perturbation.
In such settings, it more likely that each user $i$ observes a random sequence $(A^i_n)_{n\geq 1}$, 
where $A^1_n,\dots,A^N_n$ typically correspond to the realization at time $n$ of a time-varying ergodic channel. 
The distributed optimization scheme is given by equation~(\ref{eq:algo}) where
$$
Y_{n,i} = \beta_i \nabla_\theta R_i(\theta_{n,i},A^i_n)\ .
$$
Assume for the sake of simplicity that sequence $(A^1_n,\dots,A^N_n)_n$ is i.i.d. By Theorem~\ref{the:cvc},
all users converge to a consensus on the global resource allocation parameters.
After convergence of the distributed R-M algorithm, the resource allocation parameters
achieve a Kuhn-Tucker point of the optimization problem:
\begin{equation}
\max_{\theta\in G} \sum_{i=1}^N \beta_i\bE[R_i(\theta,A^i)]
\label{eq:ergo}
\end{equation}where the expectation in the inner sum is taken w.r.t. the channel coefficients $A^i$.

\begin{figure}[h]
  \centering
  \includegraphics[width=7cm]{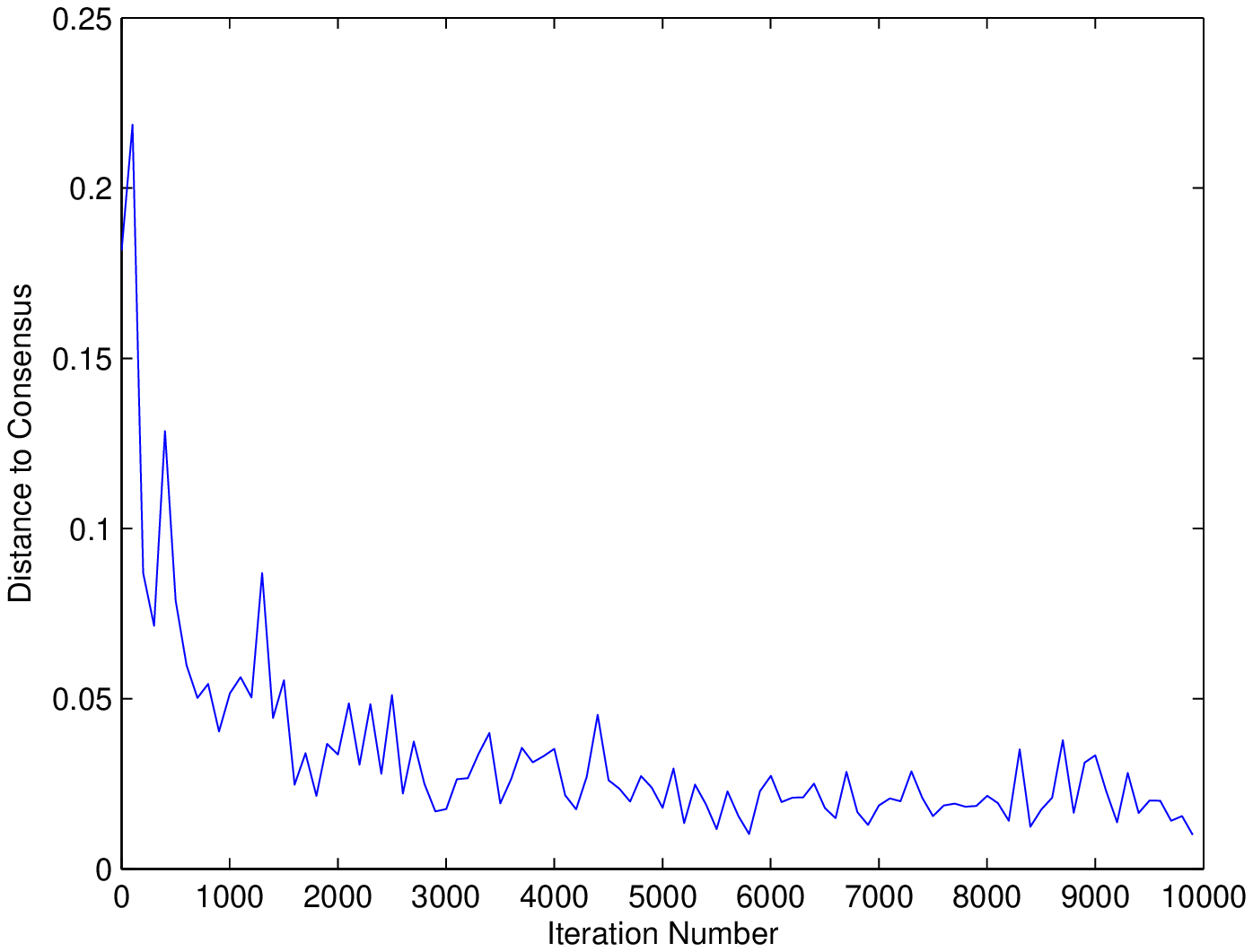}
  \caption{$|\Jo\thn|$ as a function of $n$.}
  \label{fig:agreement}
\end{figure}

\begin{figure}[h]
  \centering
  \includegraphics[width=7cm]{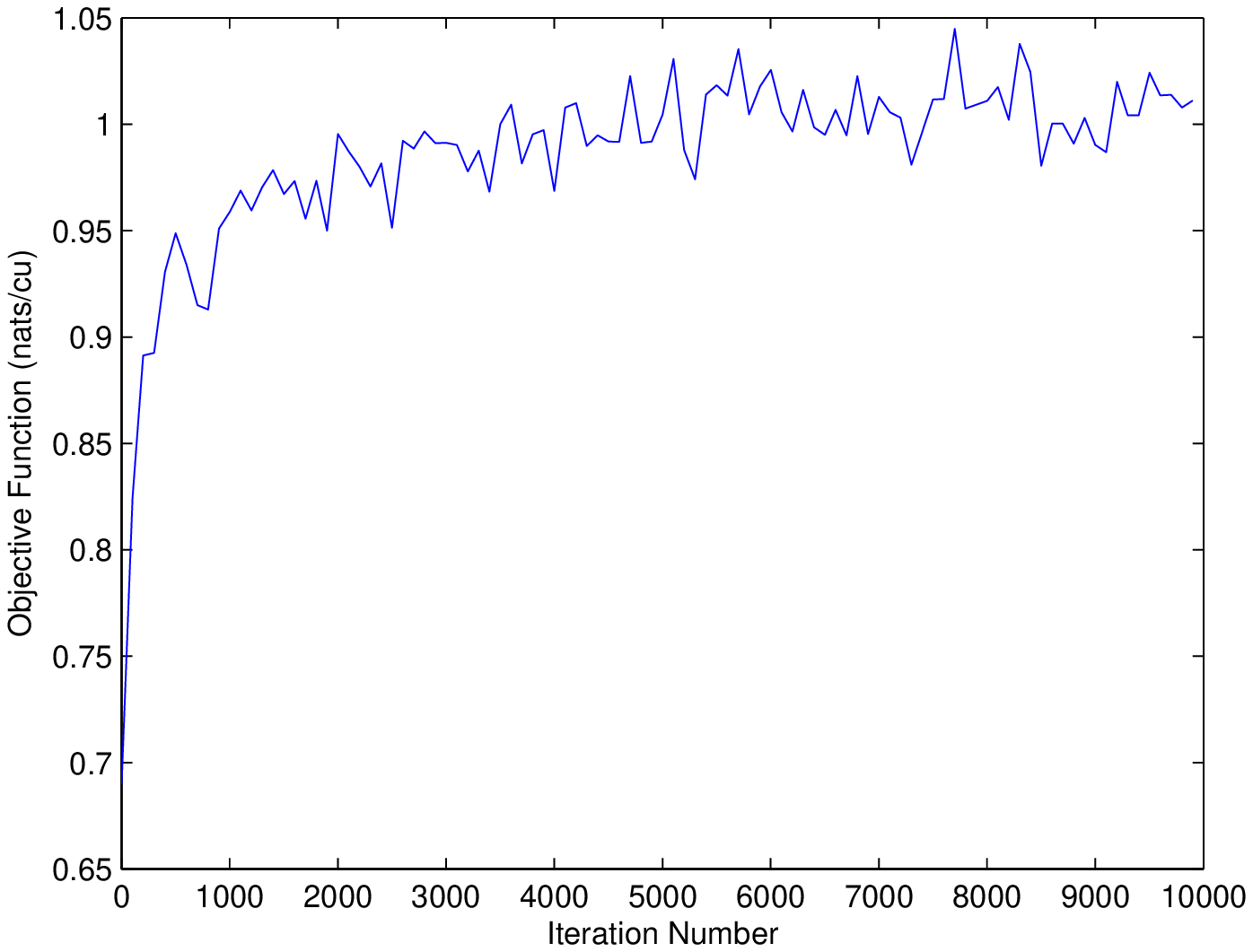}
  \caption{Estimated value of the objective function.}
  \label{fig:utility}
\end{figure}

We provide some numerical results. Consider four nodes: 1 is connected to 2 ($1\sim2$), $1\sim3$,
$2\sim3$, $2\sim 4$, $3\sim 4$. Assume $Q=2$, $\beta_1=\beta_3=0.3$, $\beta_2=\beta_4=0.2$, 
$\sigma^2_1=\sigma^2_4=0.1$, $\sigma^2_2=0.05$, $\sigma^2_3=0.02$. Assume that all r.v. $A^{i,j;k}_n$ are i.i.d.
with standard exponential distribution. 
The network model is chosen as in Section~\ref{subsec:network}.
The algorithm is initialized at a random point $\bth_0$. 
Figure~\ref{fig:agreement} illustrates the fact that the disagreement between agents $|\Jo\thn|$ 
converges to zero as $n$ tends to infinity. Figure~\ref{fig:utility} shows the estimated value
of the objective function given by~(\ref{eq:ergo}). The expectation in (\ref{eq:ergo})
is estimated using $10^3$ Monte-Carlo trials at each iteration.

\subsection*{Acknowledgements}

The Authors would like to thank Gersende Fort and Walid Hachem 
for their contribution to this work.
They are as well grateful to Eric Moulines for fruitful discussions.

\small
\bibliographystyle{IEEEbib} 
\bibliography{biblio}

\balancecolumns
\end{document}